\documentclass[a4paper]{article}

\usepackage[english]{babel}
\usepackage[utf8]{inputenc}
\usepackage{natbib,authblk,bbm,bm}
\usepackage{amsmath,setspace,geometry,amsfonts,amssymb}
\usepackage{graphicx}
\usepackage[mathlines]{lineno}
\usepackage{fullpage}
\usepackage{float}
\usepackage{caption}
\usepackage{subcaption}

\begin{document} 
\begin{spacing}{1.4}

\title{Analysis of animal accelerometer data using hidden Markov models}

\author{Vianey Leos-Barajas$^{\rm a}$\footnote{Corresponding author:\\ \indent e-mail: vianey@iastate.edu\\ \indent address: Snedecor Hall, Ames IA, Zip:50011 \\ \indent phone number: 515-509-5388}, 
Theoni Photopoulou$^{\rm b}$, 
Roland Langrock$^{\rm c}$,
Toby A.\ Patterson$^{\rm d}$,
Yuuki Watanabe$^{\rm e}$,
Megan Murgatroyd$^{\rm b}$ 
and  Yannis P. Papastamatiou$^{\rm f,g}$

\vspace{1em}

\small 
$^{\rm a}$Iowa State University, USA; $^{\rm b}$University of Cape Town, South Africa; $^{\rm c}$Bielefeld University, Germany; $^{\rm d}$CSIRO, Oceans and Atmosphere, Australia; $^{\rm e}$National Institute of Polar Research, Japan; $^{\rm f}$University of St Andrews, UK; $^{\rm g}$Florida International University, USA
}

\date{}
\maketitle
\vspace{-2em}

\begin{abstract} 
\noindent 1. Use of accelerometers is now widespread within animal biotelemetry as they provide a means of measuring an animal's activity in a meaningful and quantitative way where direct observation is not possible. In sequential acceleration data there is a natural dependence between observations of movement or behaviour, a fact that has been largely ignored in most analyses. \\%set the context and purpose for the work; 
2. Analyses of acceleration data where serial dependence has been explicitly modelled have largely relied on hidden Markov models (HMMs). Depending on the aim of an analysis, either a supervised or an unsupervised learning approach can be applied. Under a supervised context, an HMM is trained to classify unlabelled acceleration data into a finite set of pre-specified categories, whereas we will demonstrate how an unsupervised learning approach can be used to infer new aspects of animal behaviour. \\% indicate the approach and methods used; 
3. We will provide the details necessary to implement and assess an HMM in both the supervised and unsupervised context, and discuss the data requirements of each case. We outline two applications to marine and aerial systems (sharks and eagles) taking the unsupervised approach, which is more readily applicable to animal activity measured in the field. HMMs were used to infer the effects of temporal, atmospheric and tidal inputs on animal behaviour.\\ %outline the main results;
4. Animal accelerometer data allow ecologists to identify important correlates and drivers of animal activity (and hence behaviour). The HMM framework is well suited to deal with the main features commonly observed in accelerometer data, and can easily be extended to suit a wide range of types of animal activity data. The ability to combine direct observations of animals activity and combine it with statistical models which account for the features of accelerometer data offer a new way to quantify animal behaviour, energetic expenditure and deepen our insights into individual behaviour as a constituent of populations and ecosystems. %identify the conclusions, the wider implications and the relevance to management or policy. 
\end{abstract}
\vspace{0.5em}

\vspace{0.5em}

\section{Introduction} 

Accelerometers are becoming more prevalent in the fields of animal and human biotelemetry (\citealp{bao04}; \citealp{rav05}; \citealp{she08}; \citealp{alt10}). The potential of accelerometers lies in the fact that they provide a means of measuring activity in a meaningful and quantitative way where direct observation is not possible \citep{she08,nat12,bro13}. While these instruments are cheap and compact, recording acceleration at a high temporal resolution and in up to three dimensions quickly results in terabytes of data that present various challenges regarding transmission, storage, processing and statistical modelling.

Much of the focus in the analysis of acceleration data has been on identifying patterns in the observed waveforms that correspond to a known behaviour or movement mode. This can be achieved by employing statistical classification methods and entails observing the animal, manually assigning labels corresponding to behaviours to segments of the data, and training a model using the labelled data in order to subsequently classify remaining unlabelled data. There are many papers that have shown the effectiveness of various machine learning algorithms for classification of human acceleration data \citep{bao04,rav05,alt10,man10}. Algorithms such as support vector machines (SVM), classification trees, random forests, among others, have also recently been used for classification of animal acceleration data \citep{nat12,car14,gra15}. For example, \cite{nat12} compared the effectiveness of five machine learning algorithms to distinguish between eating, running, standing, active flight, passive flight, general preening and lying down, for griffon vultures. 

Most machine learning algorithms assume independence between individual observations. However, in sequential acceleration data there is a natural dependence between observations of movement or behaviour --- once initiated, particular animal behaviours often last for periods longer than the sampling frequency. This fact has been largely ignored in most applications of classification approaches. The studies where serial dependence has been explicitly modelled have largely relied on hidden Markov models (HMMs) \citep{war06,he07,man10,man11}.  Typically, and in common with the aforementioned machine learning approaches, in the training stage, the states of the HMM were known a priori, requiring corresponding data derived from direct observations. 

There are two main difficulties with such a supervised learning approach. First, while there has been much success in classification of human acceleration data, where training data can usually be obtained with minimal effort, this may not be feasible for some animals. Humans can easily be observed in a laboratory setting, given instructions, or monitored in more realistic settings, such as walking outdoors or in their home. In certain cases, animals can also be monitored in a laboratory setting, but movement patterns recorded in the lab from free-ranging animals may not appear exactly the same as in data collected while in more natural settings. Conversely, many behaviours can only be observed in natural settings \citep{she08,nat12,bro13}. 
 
Second, human acceleration data has commonly been used as a tool for health monitoring and other situations where the focus is on (state) prediction, as opposed to learning how external factors drive the behaviours. Classification of behaviours {\em alone}, while certainly of interest in many scenarios, may not lead to biologically interesting inference. Once the classification has been done, the task of relating these states to environmental (and other) covariates in order to identify drivers in behaviours remains. Moreover, as the classifications are not without error it is difficult to make appropriate inferential statements, propagating the state uncertainty through to the modelled effect of the covariates. These difficulties will arise in any classification method. 

In the supervised context, i.e.\ when classification is the main purpose of an analysis, we train the HMM to recognize specific behaviours. Alternatively, HMMs can also be used in an unsupervised context, i.e.\ when there are no labelled data. In an unsupervised context the states are not pre-defined to represent a specific behaviour, and instead will be allocated such that the model captures as much as possible of the marginal distribution of the observations as well as their correlation structure. If biologically meaningful response variables from the acceleration data are considered, then the HMM states will usually represent interpretable activity levels or even proxies of behavioural modes. Being data-driven the states can be as, if not more, informative in the unsupervised setting than the alternative. We can then incorporate exogenous or, where available, endogenous variable(s) of interest, to make inferential statements. HMMs and related state-switching models, in particular state-space models, have successfully been implemented to identify drivers of movement based on tracking data \citep{pat09}, and can similarly be applied in the context of accelerometer data.

In this paper we review HMM-based approaches to the analysis of animal accelerometer data. 
In Section 2 we will provide an overview of accelerometer data and connect the data processing step to the HMM-based approaches described in Section 3. In Section 4 we demonstrate the use of HMMs with real data examples from marine and aerial systems. 

\section{Accelerometer data} \label{sec:data}

% Number of axes 
Accelerometer devices measure in up to three axes, which can be described relative to the body of the animal; longitudinal, lateral and dorso-ventral. Acceleration recorded along one or two axes can be used to measure movement in parts of the body, e.g.\ the mandible \citep{suz09, nai10, iwa15}, or aspects of whole body acceleration, e.g.\ longitudinal surge in \cite{sak09}. Currently, acceleration is most commonly recorded in three axes, and to a lesser degree, in two axes \citep{bro13}, to measure locomotion. Especially for flying or swimming animals, measuring in all three axes is essential for capturing all movements performed by the animal, since movement is inherently three-dimensional. 
However, the resolution of acceleration is key to enabling identification of behaviours from the raw data.\\

\noindent \textit{Data Processing for Classification}\\

Although the observed (raw) acceleration data can be used to identify specific movements in animals, HMMs and other machine learning algorithms typically require more than that to accurately classify the unlabelled data. These methods will require appropriate features, i.e.\ summary statistics. In most cases, and as will be described here, a window size is selected and multiple features are calculated from the observations corresponding to each window. The window size can correspond to a number of observations or can also be a sliding window such that there is some percentage of overlap between consecutive windows. 

The derived features should be driven by the classes of movements that have been defined and chosen in such a way to accentuate the differences in observed acceleration measurements. For example, for each window we can compute the mean, variance, dynamic body acceleration, pitch, correlation among axes (and the list goes on). No optimal set of features exists  
although there are many commonalities between those used in applications of classification of acceleration data \citep{bao04, mar09,nat12, bro13}. For instance, \cite{nat12} used thirty-eight features in order to distinguish between eating, running, standing, active flight, passive flight, general preening and lying down, for griffon vultures, while \cite{gra15} used eight features to distinguish between standing, walking, swimming, feeding, diving and grooming of Eurasian beavers. In each case, means and variances of each of the three axes are used, as well as overall dynamic body acceleration (ODBA), the sum of dynamic body acceleration from the three axis, among others. \\ 

\noindent \textit{Connecting Measures to Behaviours}\\

When the aim is to classify the acceleration data, data processing is driven by identifying a set of features that can be used to distinguish between specific movements, even if those features are not themselves interpretable as a specific behaviour when considered on their own. However, there are metrics derived from the accelerometer data that, on their own, can be used as proxies for behaviour and as input to an HMM.

There can be a number of diverse signals corresponding to different movement behaviours. Repeating patterns in at least one axis tend to arise from regular movement such as stroking \citep{sak09}, flapping, running or walking \citep{she08}, whereas sudden changes, corresponding to bursts of acceleration, are often associated with prey pursuits or capture \citep{suz09, sim12, yde14, hee14}, as well as predator avoidance, or conflict. 

% Useful metrics derived from acceleration data
In addition to behaviour, several measures can be used to summarise effort or exertion and relate acceleration to energy expenditure or activity levels. These include overall dynamic body acceleration (ODBA) \citep{wil06, gle11, ell13, gle13} and vectorial dynamic body acceleration (VeDBA) \citep{qas12}, while minimum specific acceleration (MSA) \citep{sim12} can be used to disentangle the gravitational component of acceleration (static acceleration) from the movement signal, or specific acceleration (also dynamic acceleration). One of the simplest and most unambiguous interpretations of static acceleration data is body posture, which in many cases can be directly interpreted as a specific behaviour \citep{wil08, she08}. 

Speed is an important aspect of animal movement since it is often related to energy expenditure or energy acquisition. Speed can be calculated using acceleration and depth data from marine animals during diving, when the pitch angle is steep \citep{wat03, she08}. Stroke frequency in aquatic animals such as sharks \citep{wat12}, whales, seals \citep{sat03}, and diving seabirds \citep{sat08} can be used both as a measure of activity and, in some cases, a measure of foraging success \citep{sat08}. It has also been shown that acceleration can be used to distinguish between different flight behaviours in soaring birds \citep{wil15}.

\section{Analysis of accelerometer data} \label{sec:analysis}

We will first provide a brief overview of the HMM framework (Section \ref{subsec:HMM}). Subsequently, in Section \ref{subsec:HMMclass}, we will review how HMMs can be used for classification of animal accelerometer data. In Section \ref{ss:covinf}, we focus on the implementation of HMMs in an unsupervised setting, such that the meaning of the states is driven entirely by the data and the focus lies on general inference rather than classification only. 

\subsection{Hidden Markov models} \label{subsec:HMM}

An HMM is a stochastic time series model involving two layers: an observable {\em state-dependent process},  denoted by $\{ Y_t \}_{t=1}^T$ (in the univariate case), and an unobservable {\em state process}, denoted by $\{ C_t \}_{t=1}^T$. 
The state-dependent process models the observations, while the state process is a latent factor influencing the distribution of the observations. In our case, the observations are the accelerometer metrics considered, and the latent states are closely related to the animal's behavioural state. More specifically, the state process $\{ C_t \}$ takes on a finite number of possible values, $1,\ldots,M$, and its value at time $t$, $c_t$, selects which of $M$ possible component distributions generates observation $y_t$. The Markov property is assumed for $\{ C_t \}$, such that evolution of the process over time is completely characterized by the one-step state transition probabilities.

These models are natural and intuitive candidates for modelling animal accelerometer data, for two reasons: 1) they directly account for the fact that any corresponding observation will be driven by the underlying behavioural state, or general activity level, of the animal, and 2) they accommodate serial correlation in the time series by allowing states to be persistent. HMMs seek to capture the strong autocorrelation in accelerometer data in a mechanistic way, rather than either neglecting this feature completely or only including it in a nuisance error term. HMMs can therefore be used for inference on complex temporal patterns, including the behavioural state-switching dynamics and how these are driven by environmental variables.

To complete the basic HMM formulation, we first summarize the probabilities of transitions between the different states in the $M \times M$ transition probability matrix (t.p.m.) $\boldsymbol{\Gamma}=\left( \gamma_{ij} \right)$, where $\gamma_{ij}=\Pr \bigl(C_{t}=j\vert C_{t-1}=i \bigr)$ (for any $t$), $i,j=1,\ldots,M$. Note that here we are assuming that the state transition probabilities are constant over time; this assumption will be relaxed in Section \ref{ss:covinf}. The initial state probabilities are summarized in the row vector $\boldsymbol{\delta}$, where $\delta_{i} = \Pr (C_1=i)$, $i=1,\ldots,M$. 

Second, we need to specify state-dependent distributions (sometimes called emission distributions), $p(y_t | C_t = m)$, or more succinctly $p_m(y_t)$, for $m=1,...,M$. These distributions can be discrete or continuous, and possibly also multivariate (in which case we write $\mathbf{y}_t=(y_{1t},\ldots,y_{Rt})$). Usually, the same parametric distribution is assigned to all $M$ states, such that each state differs in terms of its associated values of the parameters. Selection is driven by the data itself, e.g.\ count data or continuous observations. 

\subsection{Classification based on supervised learning} \label{subsec:HMMclass}

HMMs provide a solid framework for the classification of data with strong serial dependence, such as sequential acceleration data, which are often processed to represent movements over a few seconds, or less, at a time \citep{war06, he07, man10}. In this section, we will cover the implementation and testing of an HMM when there are labelled data available and the aim is to categorize unlabelled data into a finite set of pre-defined classes, with a low error rate. 

In supervised learning, the collection of labelled time series is split into three categories: training data, validation data, and testing data. An approximate split of 50$\%$, 25$\%$, 25$\%$ can be used \citep{has01}, although this will require splitting the observations in such a way that takes into account the temporal dependence and possible multiple time series from the same animal. The training data will be used to fit the HMM, the validation data is used for model selection, and the testing data will be used to estimate the prediction error for the final model selected. We can test a variety of HMM formulations and select the best performing among them. 

However, in many cases there will only be sufficient data to split into training and testing, in which case multiple approaches can be taken for model selection/assessment, as described in \cite{has01}. Cross-validation is one of the most common approaches to estimate prediction error. If the $K$ time series are independent of each other, we can use leave-one-out cross-validation treating each time series as an observation. We fit the HMM to $K-1$ time series, the training data, and decode the state sequence of the time series left out, the testing data, using the parameter estimates of the fitted model, in order to compare our predictions to the known state sequence. This process is repeated $K$ times. (Leave-one-out cross-validation is one approach that can be taken, but certainly not the only one, and is only used here for illustrative purposes.)

In the training stage, the state sequence is known which greatly simplifies fitting the HMM. If the sequence of states, $c_1,\ldots,c_T$, is known, then we obtain the (complete-data) likelihood of the HMM for one time series as
\begin{equation} \label{eq:compdatlik}
\it{L}_C = p(c_1\ldots,c_T,\mathbf{y}_1,\ldots,\mathbf{y}_T) = \delta_{c_1} p_{c_1}(\mathbf{y}_1)\prod_{t=2}^T \gamma_{c_{t-1}, c_{t}} p_{c_t}(\mathbf{y}_t).
\end{equation}
For $K$ independent time series, the likelihood is the product of the individual complete-data likelihoods. For an HMM in the context of classification, each state-dependent distribution is often selected to be a multivariate normal distribution (MVN), or a mixture of MVNs if the data are highly non-normal. The MVN is commonly used as it can model the dependence between random variables, in our case the $R$ features that will be used for classification. (Alternatively, one can assume conditional independence between all features, given the states, such that the joint density, p$(\mathbf{y}_t \vert C_t = m)$, can be written as p$(\mathbf{y}_t \vert C_t = m)$ = $\prod_{r=1}^R$p$(y_{rt} \vert C_t = m)$.)  We will use the MVN to illustrate the procedures for training an HMM for classification of unlabelled data.  
The density of the state-dependent MVN is 
\begin{equation*}  
p(\mathbf{y}_t| C_t = m) = \frac{1}{(2\pi)^{R/2} \vert \boldsymbol{\Sigma}_m \vert ^{1/2}}  \text{exp}\left(-\frac{1}{2} (\mathbf{y}_t - \boldsymbol{\mu}_m)^{\top} \boldsymbol{\Sigma}_m^{-1}(\mathbf{y}_t - \boldsymbol{\mu}_m) \right),
\end{equation*}
where $\boldsymbol{\mu}_m$ and $\boldsymbol{\Sigma}_m$ are the vector of state-dependent means (of the $R$ variables) and the state-dependent covariance matrix, respectively. These need to be estimated for each state $m$, alongside the state transition probabilities and the initial state distribution. Thus, we maximize (\ref{eq:compdatlik}) (or, in case of multiple time series, a product of likelihoods of type (\ref{eq:compdatlik})) with respect to  $\boldsymbol{\delta}$, $\boldsymbol{\Gamma}$, $\boldsymbol{\mu}_m$ and $\boldsymbol{\Sigma}_m$,  $m=1,\ldots,M$, to obtain the maximum likelihood estimates (MLEs). Since the states are known the maximization task conveniently splits into several independent parts, each of which is fairly straightforward. First, the $m$-th entry of $\hat{\boldsymbol{\delta}}$ is simply the proportion of the time series that start in state $m$. Second, the entries of the t.p.m.\ are estimated by
\begin{linenomath}
\begin{equation*}
\hat{\gamma}_{ij} = \frac{\# \text{ transitions from state } i \text{ to state } j}{\text{total } \# \text{ transitions from state } i},
\end{equation*}
\end{linenomath}
for $i,j = 1,...,M$. (Note this is the MLE conditional on the initial state, $c_1$.)
Finally, $\boldsymbol{\mu}_m$ and $\boldsymbol{\Sigma}_m$ are estimated using only the observations allocated to state $m$. 

Given a fitted HMM,  two state-assignment approaches can be used in the testing stage: local or global decoding. Local decoding assigns a state at each time $t$ by maximising the conditional probability Pr$(C_t \vert \mathbf{Y}_1, \ldots, \mathbf{Y}_T)$, under the fitted model. Global decoding maximises $\text{Pr}\left(C_1, \ldots, C_T \vert \mathbf{Y}_1, \ldots, \mathbf{Y}_T \right)$ to assign the most likely sequence of states for the entire time series. A dynamic programming algorithm, the Viterbi algorithm, can be used to find the optimal state sequence at low computational effort. While the outcome is usually very similar, it does sometimes happen that the decoded states differ. Essentially Viterbi considers the state sequence as a whole, while local decoding considers each time point in isolation. Full details are provided in \cite{zuc09}. The state predictions are compared to the known states, and the proportion of correctly decoded states serves as an estimate of the prediction accuracy.

\subsection{Inference using unsupervised learning} \label{ss:covinf}

So far, we have been focusing on the case where there is a training sample, i.e.\ acceleration data together with the associated behavioural states. Corresponding analyses involve training the HMM based on such labelled data and then using that HMM to categorize incoming new, unlabelled data. While certainly of interest in some settings, in practice, more often than not labelled data will not be available, but only the accelerometer data. In such unsupervised settings, the HMM framework can be equally useful, but is typically applied for different purposes than in classification. More specifically, the meaning of the states in such cases is often not of interest {\it per se}. Instead, an HMM is used simply as an approximate representation of the real data-generating process, and this may or may not entail that the nominal HMM states are biologically meaningful. (However, metrics derived from the accelerometer data, as described in Section \ref{sec:data}, have been shown to provide insight into activity levels or correspond to classes of behaviours, such that when used as response variables in the HMM these can lead to biologically interpretable states.) Unsupervised learning of HMMs for accelerometer data has the advantage that the states are estimated in a data-driven manner. In particular, for many of the metrics described in Section \ref{sec:data} that are connected to behaviours, assignment of classes is difficult, to say the least, especially for animals where behaviours are not well-defined. These include animals which cannot be directly observed for long periods such as aquatic organisms. In these cases HMMs have been shown to be useful. For example, \cite{phi15} applied HMMs using MVN responses of behavioural indices but in an unsupervised context to understand the behaviour of free swimming tuna from vertical movement data collected by data-storage tags. Below we demonstrate the application of unsupervised learning for another difficult to observe species, namely blacktip reef sharks.  

There are three different possible purposes of having an approximate representation of the real process: (i) a description of the stochastic structure of the system; (ii) extraction of information; (iii) prediction \citep{kon08}. In the ecological literature on animal movement modelling, HMMs are used primarily to address (i) and (ii), the former in the sense that concise descriptions of movement patterns are sought, the latter in the sense that inference on the interaction of animals with their environment is drawn. In general, the ability to make inferential statements provides an avenue to answer questions about the behavioural processes, movement patterns and transitions between behaviours under different conditions or in relation to other covariates. 

Addressing a research question related to aim (ii) usually involves the incorporation of covariates into the statistical model. In the HMM setting, this is commonly done at the level of the hidden states. For the general case of time-varying covariates, we define the corresponding time-dependent transition probability matrix $\bm\Gamma^{(t)} = (\gamma_{ij}^{(t)})$, where $\gamma_{ij}^{(t)} = \Pr(C_{t+1}=j \vert C_t=i)$. The transition probabilities at time $t$, $\gamma_{ij}^{(t)}$, can then be related to a vector of environmental (or other) covariates, $\bigl(\omega_1^{(t)},\dots,\omega_p^{(t)}\bigr)$, via the multinomial logit link:
\begin{linenomath}
\begin{equation*}
\gamma_{ij}^{(t)} = \dfrac{\exp(\eta_{ij})}{\sum_{k=1}^N \exp(\eta_{ik})}, \qquad \text{where} \qquad
\eta_{ij} = 
\begin{cases}
	\beta_0^{(ij)} + \sum_{l=1}^p \beta_l^{(ij)} \omega_l^{(t)} & \text{ if } i \neq j;\\
	0 & \text{ otherwise}.
\end{cases}
\end{equation*}
\end{linenomath}
Essentially there is one multinomial logit link specification for each row of the matrix $\bm\Gamma^{(t)}$, and the entries on the diagonal of the matrix serve as reference categories. 

While with labelled data the likelihood of interest is the complete-data likelihood given in (\ref{eq:compdatlik}) (i.e.\ the joint density of observations and states), for unlabelled data the likelihood of interest is the density of the observations only, $\it{L} = p(\mathbf{y}_1,\ldots,\mathbf{y}_T)$, the evaluation of which requires the consideration of all possible state sequences that might have given rise to these data. Given the simplicity of the complete-data likelihood, it comes as no surprise that the expectation-maximization (EM) algorithm is seen as a natural approach to deal with this complexity. The EM algorithm is an iterative scheme for finding the MLEs, and involves alternate updates of the conditional expectation of the states (given the data and the current estimates) and updates of the model parameters based on the complete-data log-likelihood where the unknown states are replaced by their conditional expectations. Despite the intuitive appeal of EM, as pointed out for example by \citet{mac14}, there is a more direct route to MLEs, namely direct numerical maximization of the likelihood. Since it is our view that users are better off focusing on the simpler direct maximization approach, it is only this approach that we present here in detail. 

In order to efficiently calculate (and numerically maximize) the likelihood, rather than separately considering all $M^T$ possible hidden state sequences that might have given rise to the observations, which in general is infeasible, the HMM dependence structure can be exploited to perform the likelihood calculation recursively, traversing along the time series and updating the likelihood and state probabilities at every step. To do so, we define the forward probability of state $j$ at time $t$ as $ {\alpha}_{t}(j) = p(\mathbf{y}_1,...,\mathbf{y}_t,c_{t}=j)$. Due to the dependence structure, we have
\begin{linenomath}
\begin{align*}
{\alpha}_{t}(j) 
& = p(\mathbf{y}_t \mid \mathbf{y}_1,..., \mathbf{y}_{t-1},c_{t}=j) p(\mathbf{y}_1,..., \mathbf{y}_{t-1},c_{t}=j) \\
& = p(\mathbf{y}_t \mid c_{t}=j) \sum_{i=1}^M  p(\mathbf{y}_1,..., \mathbf{y}_{t-1},c_{t}=j,c_{t-1}=i) \\
& = p(\mathbf{y}_t \mid c_{t}=j) \sum_{i=1}^M  p(c_{t}=j \mid \mathbf{y}_1,..., \mathbf{y}_{t-1},c_{t-1}=i) p(\mathbf{y}_1,..., \mathbf{y}_{t-1},c_{t-1}=i) \\ 
& = \sum_{i=1}^M  {\alpha}_{t-1}(i) \gamma_{ij}^{(t-1)}  p_j(\mathbf{y}_t) .
\end{align*}
\end{linenomath}
In matrix notation, defining the vector of forward probabilities at time $t$ as 
$\boldsymbol{\alpha}_{t} = \bigl( {\alpha}_{t}(1), \ldots, {\alpha}_{t}(M) \bigr)$, this becomes $ \boldsymbol{\alpha}_{t} = \boldsymbol{\alpha}_{t-1} \boldsymbol{\Gamma}^{(t-1)} \mathbf{Q}(\mathbf{y}_{t})$,
where $\mathbf{Q}(\mathbf{y}_{t})= \text{diag} \bigl( p_1 (\mathbf{y}_{t}), \ldots, p_M (\mathbf{y}_{t}) \big)$. Together with the initial calculation
$ \boldsymbol{\alpha}_{1} = \boldsymbol{\delta} \mathbf{Q}(\mathbf{y}_{1})$, 
this is the so-called forward algorithm. The forward algorithm can be applied in order to first calculate $\boldsymbol{\alpha}_{1}$, then $\boldsymbol{\alpha}_{2}$, etc., until one arrives at $\boldsymbol{\alpha}_{T}$. The likelihood then simply is $\it{L}=\sum_{m=1}^M {\alpha}_{T}(m)$. The computational cost of evaluating the likelihood is linear in the number of observations, $T$, such that a numerical maximization of the likelihood becomes feasible in most cases. The structure of the likelihood evaluation is independent of the type of state-dependent distributions applied. In other words, changing the class of distributions used for the observations leads to minimal changes in the code, since only the matrices $ \mathbf{Q}(\mathbf{y}_{t})$ need to be modified. The same holds for the specification of the transition probabilities (with or without dependence on covariates). 

Model selection techniques, in particular information criteria, can be used to choose an adequate family of state-dependent distributions, to select an appropriate number of states, or to determine whether or not a covariate should be included in the model. However, users should not blindly follow such information criteria, especially with regard to the selection of the number of states. For animal behaviour data, in our experience such formal model selection approaches tend to favour overly complex models, often to an extent such that selected models become near-impossible to interpret and very difficult to work with in practice \citep{lan15}. One explanation for this is that, while clearly natural and appealing, HMM-type models are usually still very simplistic relative to the actual behavioural decisions and hence the data-generating process. A model selection scheme might then point to a model where minor, possibly irrelevant features of the data that are not well described by say a simple two-state or three-state model are captured by additional states, often such that these are rarely ever visited by the animal considered. While corresponding models might of course better explain the data at hand, it often turns out that meaningful inference is severely hindered by models with complex state architectures. In such cases a healthy dose of pragmatism is required. Applying nonparametric state-dependent distributions, as suggested by \citet{lan15}, is one strategy for avoiding overly complex state processes. 

The HMM framework encompasses various other useful tools for drawing inference, including, {\it inter alia}, the option to incorporate random effects to account for individual heterogeneity \citep{mck15} and (pseudo-)residuals for comprehensive model checks \citep{zuc09}. Furthermore, the dependence structure can be modified in various ways, e.g.\ allowing for more complex memory in the state process \citep{lan12a} without losing the ability to efficiently calculate the likelihood.

\section{Real data examples} \label{sec:dataexam}

\subsection{Modelling activity in a soaring raptor} 

Large soaring birds like raptors rely heavily on meteorological conditions and underlying topography for generation of the updrafts necessary for low-energy flight. However, the relationships between different weather conditions and bird activity patterns are not well understood. In addition, the same conditions that are favourable for soaring flight are often also favourable for the generation of renewable energy from wind. Rotating wind turbines pose a threat to many bird species, so understanding the conditions under which birds spend most time active can be important at prospective installation sites \citep{kat13, rus14}. Acceleration data, in combination with other types of movement data, hold promise in distinguishing between different kinds of flight behaviour and how active each flight behaviour is, in soaring birds \citep{wil15}.

\begin{figure}[!htb]
\centering
\includegraphics[width=\textwidth]{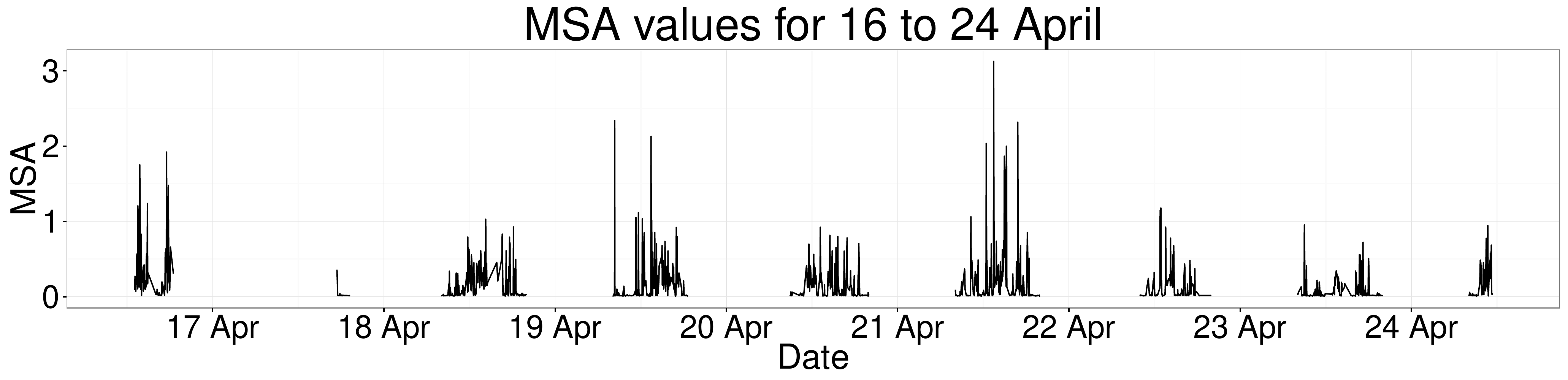}
\caption{Minimum specific acceleration values derived from three-axis acceleration data from one eagle collected over 9 days, 16-24 July 2013.}
\label{fig:msa_obs}
\end{figure}

Acceleration data were collected for an adult Verreaux's eagle {\em (Aquila verreauxii)} in the Western Cape, South Africa, in 2013, instrumented with a remotely downloadable multi-sensor data-logger (UvABiTS, University of Amsterdam, The Netherlands). Data were collected over a one-second sampling period at a resolution of 20Hz and at intervals of 112sec, for 9 days (Figure \ref{fig:msa_obs}). Each day contained a variable number of acceleration samples. Because of the discontinuous nature of the data, we used minimum specific acceleration (MSA) to extract an index of activity from the three-axis acceleration data. We were not interested in the pattern of MSA within the one-second sample so we took the maximum value from each one-second sample. This resulted in 1429 observations, with a mean number of 158.78 observations per day (SE 26.67), which equates to an observation period of approximately 5 hrs per day, on average. Data were not collected at night which created gaps between daytime bouts of observations. 

No concurrent direct observations were available, so it was not possible to link distinct behaviours with different values of MSA. Instead, we used MSA as a proxy for activity and fitted a 2-state HMM to investigate the effect of wind speed and temperature on bird activity. Without considering any ancillary data, it seemed sensible to assume two states, one including low-level activity, and the other more active movement. Preliminary data analysis revealed that MSA showed two peaks close to zero, suggesting that two types of movement comprise the less active state. We chose to formulate a model that would lump these two types of activity levels together in one state, since the interest lies in identifying higher-level energetic activity, such as flapping and hunting, instead of discriminating between, for example, being completely stationary (roosting) and preening or being otherwise alert. Therefore, we implemented a mixture of two Gamma distributions as the state-dependent probability distribution for the less active state to account for these two processes, and a single Gamma distribution in the more active state.

Since there were long gaps between the data collected on different days (due to the absence of any night-time observations), we treated the observations from each day as a separate, independent time series. We fitted the model jointly to all maximum MSA values, collected on the nine days of observation, and estimated global parameters across all segments. 

\begin{figure}[!htb] 
\centering
\includegraphics[scale=.35]{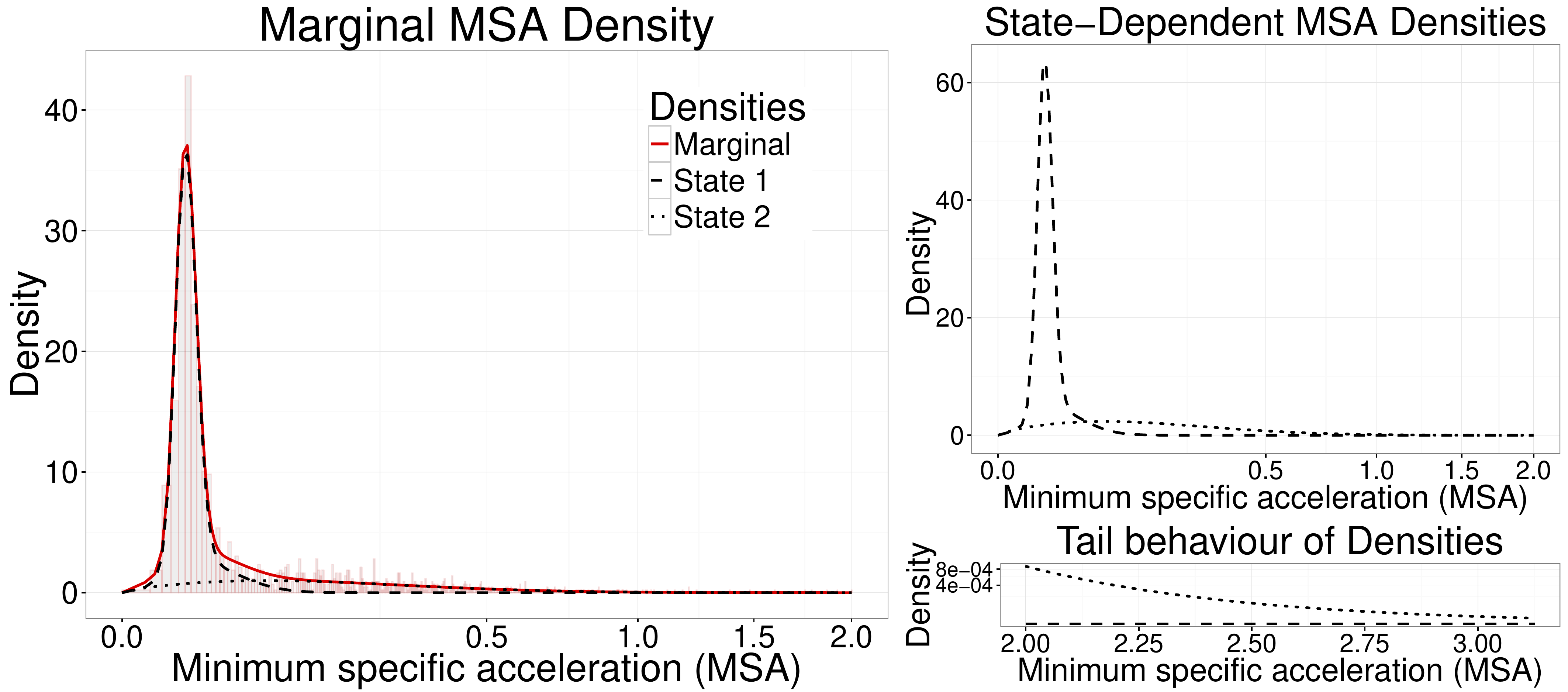}
\caption{Histogram of minimum specific acceleration, truncated at MSA=2, with marginal density and state-dependent densities weighted according to the proportion of observations assigned to each state. (left). Unweighted state-dependent densities (top right) and close-up of the tail behaviour of the densities (bottom right). A square root coordinate transformation for the x-axis was used in all plots and for the y-axis only for the tail behaviour plot.}
\label{fig:state_dep_msa_densities}
\end{figure}

Lift availability is known to be driven largely by wind speed and temperature, as well as their interaction with the underlying topography, though other factors also contribute. Lift adequate for soaring flight is generated by two mechanisms; (1) by upward thermal convection of air warmed by solar radiation \citep{ako10} (thermal soaring), and (2) by the movement of air over slopes and ridges in the landscape (orographic or ridge soaring). Activity levels in raptors are expected to vary throughout the day but we chose not to include time of day as a covariate for the t.p.m. because what drives this pattern is most likely the environmental variables, such as the building heat of the day and associated convection, rather than the time of day itself (though prey availability is also an important factor not considered here). April is autumn in Southern Africa which can be very warm in the day, but is still cool overnight and is characterised by fairly light winds relative to summer. The range of temperatures and wind speeds experienced by the bird during the study period ranged between 12.3 and 31.5 $^{\circ}$C, and 0 and 7.4 m/sec. 
In this example we only examine the effects of meteorological variables on the t.p.m. To capture the effect of wind and solar radiation on the probability of transitioning between states, we included wind speed, temperature and their interaction in the model as covariates affecting the entries of the t.p.m. The model including wind speed alone was favoured by the Bayesian Information Criterion (BIC)  and the Akaike Information Criterion (AIC) (Table \ref{tab:eagle_results}). The fitted state-dependent densities are shown in Figure \ref{fig:state_dep_msa_densities}.

\begin{table}[!htb]
\centering
\begin{tabular}{l c c c c c}
\hline
Model & Log-likelihood & AIC & $\Delta$ AIC & BIC & $\Delta$ BIC \\
\hline \hline
No covars & 2344.8 & -4665.6 & 23.7 & -4602.4 & 23.1 \\
Temperature & 2347.4 & -4670.9 & 17.8 & -4607.7 & 17.8 \\
\textbf{Wind speed} & 2356.4 & -4688.7 & \textbf{0} & -4625.5 & \textbf{0} \\
Wind speed, Temperature & 2356.4 & -4688.1 & 0.6 & -4614.4 & 11.1 \\
Wind speed, Temperature, & 2358.5 & -4684.9 & 3.8 & -4600.7 & 24.8 \\
\hspace{2em} Wind speed * Temperature &  &  & \\
\hline
\end{tabular}
\caption{Model fitting results for a Verreaux's eagle. Based on the AIC and BIC, the model selected includes only wind speed.}
\label{tab:eagle_results}
\end{table}

\begin{figure}[!htb]
\centering	
\begin{minipage}{.45\textwidth}
  \centering  
  \includegraphics[width=.9\linewidth]{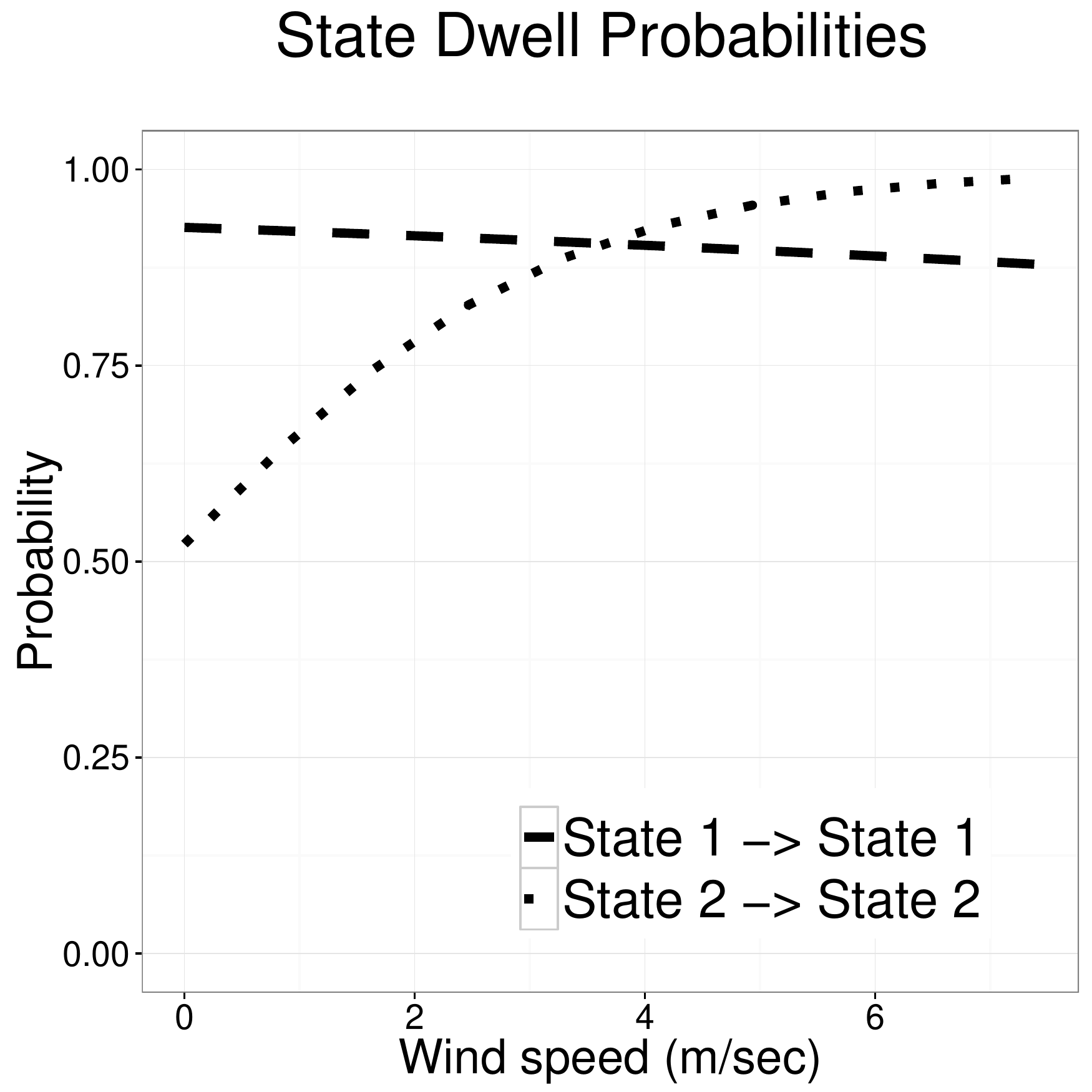}
\end{minipage}
\begin{minipage}{.45\textwidth}
  \centering
  \includegraphics[width=.9\linewidth]{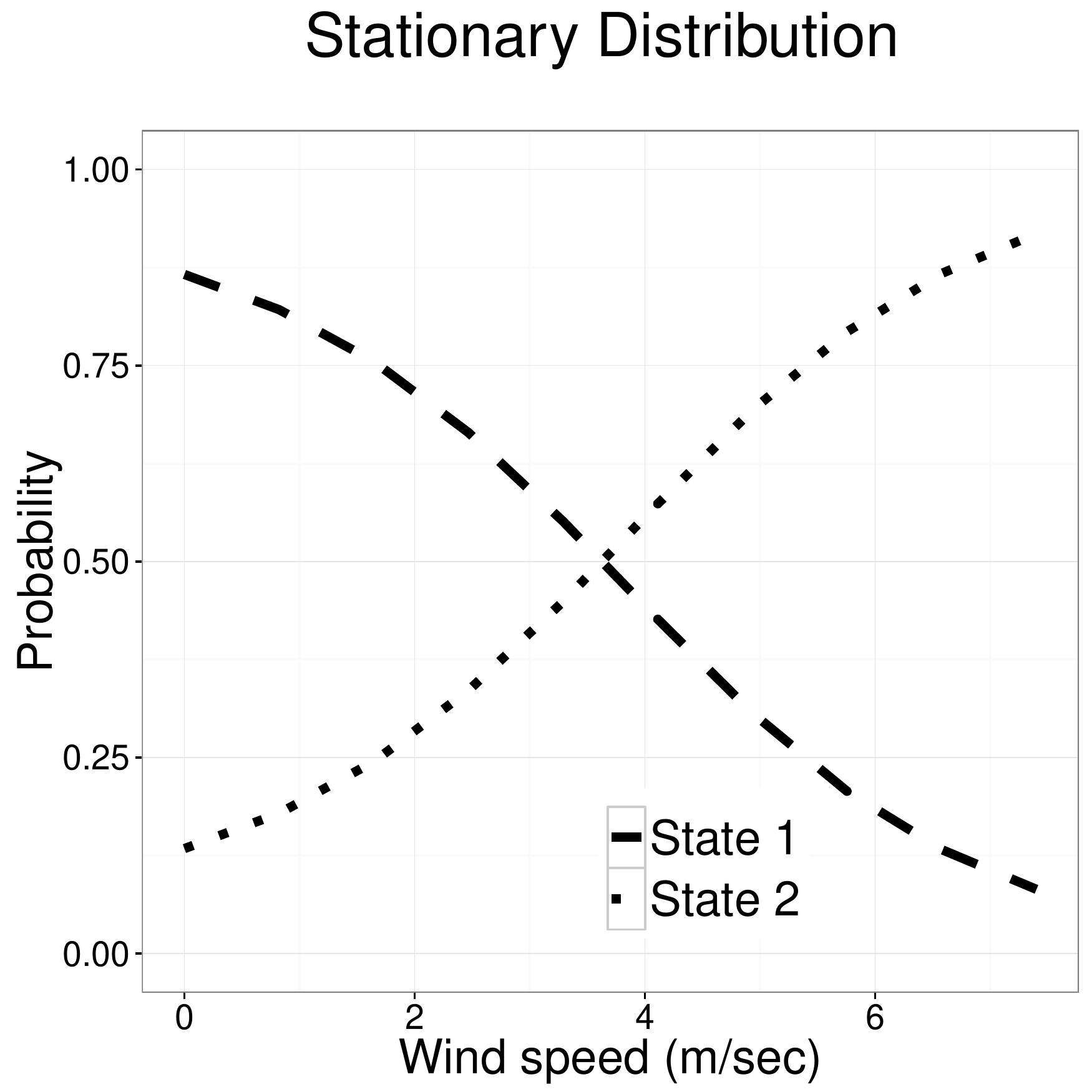}
\end{minipage}
\caption{Estimated state-dwell probabilities as a function of wind speed (left), and estimated equilibrium state probabilities as a function of wind speed (right).}
\label{fig:comp_fig_eagle}
\end{figure}

The estimated state transition probabilities suggest that, as wind speed increases, (i) the eagle has a very slightly increased chance of switching to the high-activity state when in the low-activity state, and (ii) generally spends much longer periods of time, on average, in the active state. As a consequence, the equilibrium (stationary) distribution for fixed wind speeds \citep{pat09} indicates that the bird spends more time in the active state overall as wind speed increases (Figure \ref{fig:comp_fig_eagle}). If the active state is representing orographic soaring, this would be consistent with the results of studies on migrating golden eagles \textit{Aquila chrysaetos} which also did more orographic soaring as a response to higher wind speed \cite{lan12}. The fact that temperature was not found to have a strong effect is somewhat surprising since one would expect high temperatures to be conducive to the formation of thermals and thermal soaring, and the appearance of a signal in the MSA data pointing towards activity. However, it may be that thermal soaring is a low-effort behaviour, compared to orographic soaring, and does not result in a peak in activity, as measured by the MSA. Alternatively, perhaps there were other factors in play during the study period, which made thermal soaring a less cost-effective mode of transport and the bird simply did not engage in that behaviour consistently.

\subsection{Diel activity changes in a reef-associated shark}

Many species of shark are upper trophic level predators may serve an important role in marine ecosystems. However determining the intensity of their predatory behaviour requires modelling the temporal component as their activity levels are likely to follow a diel and/or tidal cycle \citep[e.g.][]{gle13,pap15}.  Acceleration sensors provide a direct measure of activity, however, many species of shark swim continuously making it difficult to define specific behaviours (e.g.\ they are never truly at rest), making conventional classification methods problematic. HMMs can identify changes in behavioural states and how these may be related to time of day, tidal state, swimming depth, or water temperature.  To demonstrate this, we applied HMMs to accelerometry data obtained from a free-ranging blacktip reef shark (\textit{Carcharhinus melanopterus}) at Palmyra Atoll in the central Pacific ocean \citep[data taken from][]{pap15}. A multi-sensor package was attached to the dorsal fin of a 117 cm female shark. The multi-sensor data-logger (ORI400-D3GT, Little Leonardo, Tokyo, Japan) recorded 3D acceleration (at 20Hz), depth and water temperature (at 1Hz) and was embedded in a foam float which detached from the animal after four days \citep[see][]{pap15}. The package also contained a VHF transmitter allowing recovery at the surface after detachment.

\begin{figure}[!htb] 
\centering
\includegraphics[scale=.45]{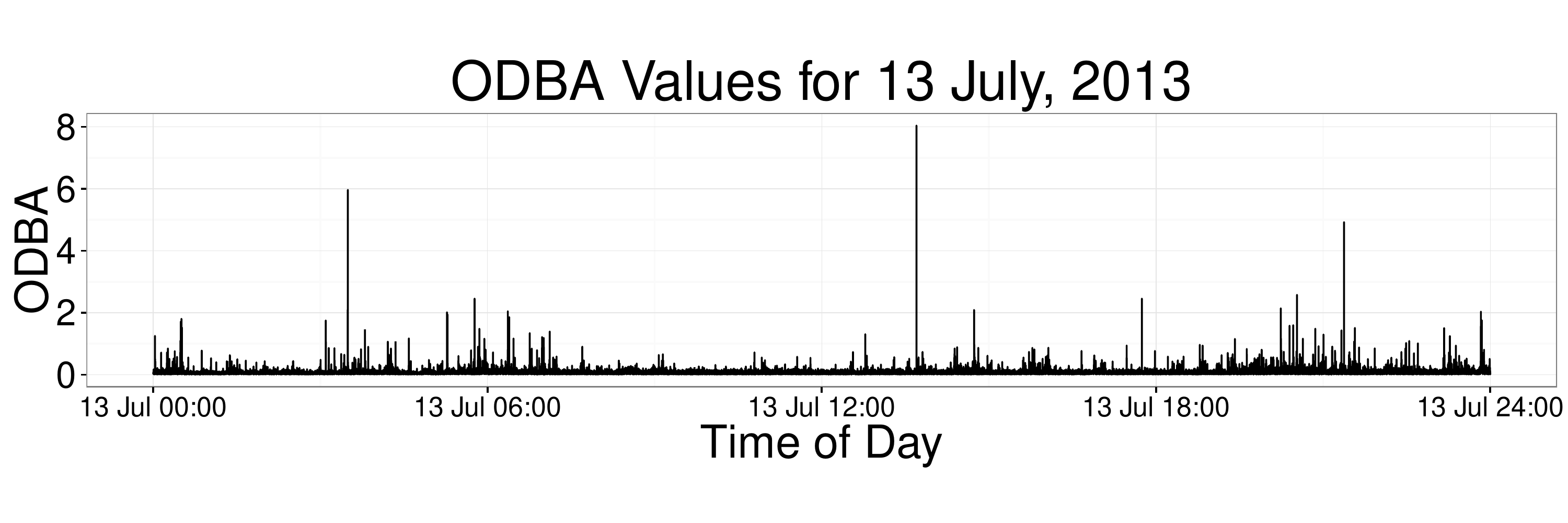}
\caption{Overall dynamic body acceleration values averaged over 1-second intervals for 13 July, 2013.}
\label{fig:odbasamp}
\end{figure}

In order to examine active behaviour, we calculated the average overall dynamic body acceleration (ODBA) of the shark over one second intervals, which resulted in 321,815 observations (after removing the first four hours of data). Figure \ref{fig:odbasamp} displays the ODBA time series of one day. Compared to metrics such as tail-beat frequency, ODBA has the advantage of measuring change in behaviour in all axes. For example, if the shark is nose down at the seafloor, attempting to capture prey, its tail-beat frequency may be low but it is still active. As we are interested in the times of day the shark was more active, as well as tide effects, we applied a 2-state HMM with one state for less active behaviour (since the shark is never completely at rest) and another for more active behaviour.

\begin{figure}[!htb] 
\centering
\includegraphics[scale=.33]{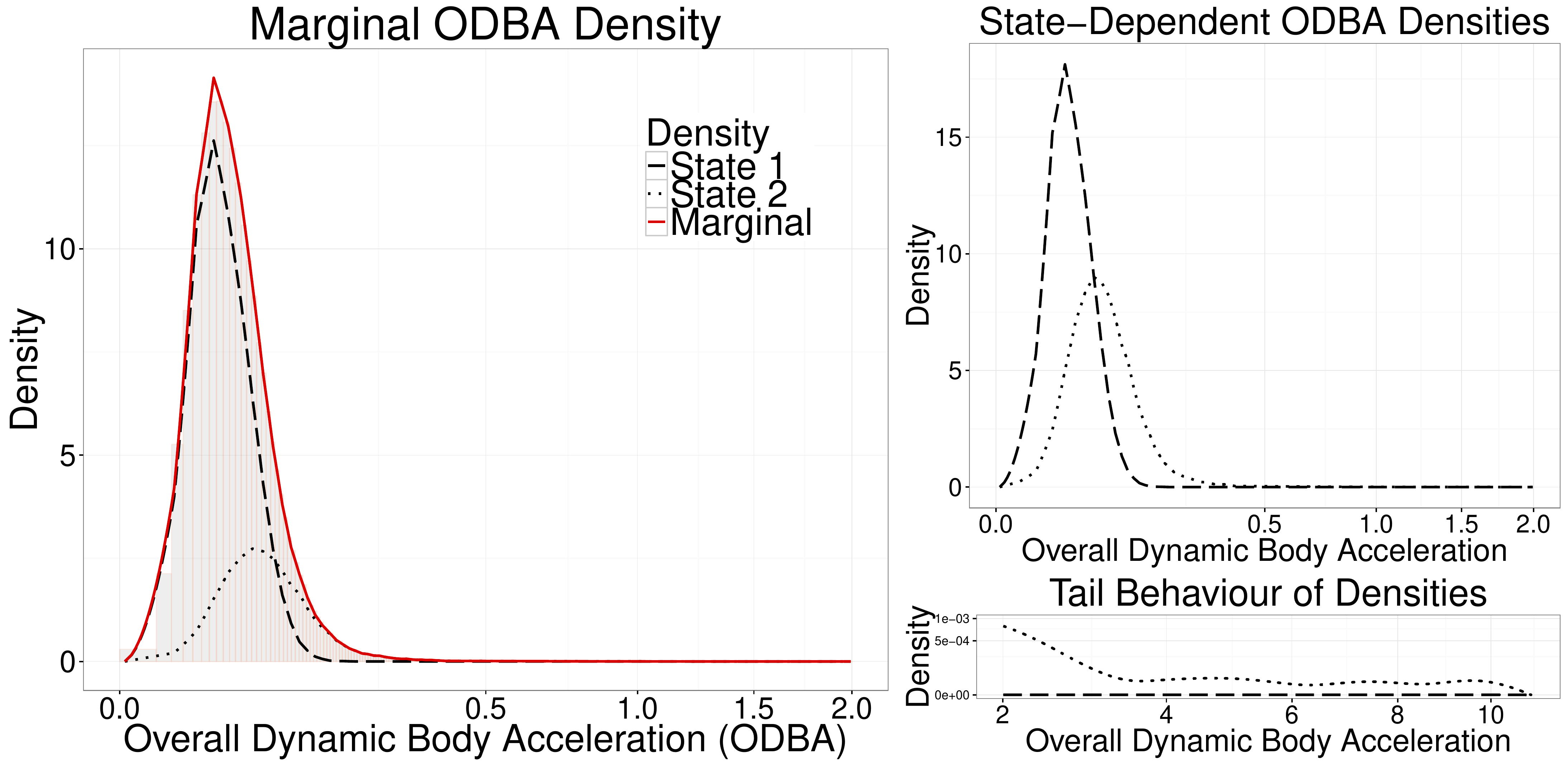}
\caption{Histogram of overall dynamic body acceleration, truncated at ODBA=2, with marginal density and state-dependent densities weighted according to the proportion of observations assigned to each state. (left). Unweighted state-dependent densities (top right) and close-up of the tail behaviour of the densities (bottom right). A square root coordinate transformation for the x-axis was used in all plots and for the y-axis only for the tail behaviour plot. }
\label{fig:statedepdens}
\end{figure}

Although there are clear spikes in ODBA that point to higher energetic activities, various combinations of parametric distributions for state 1 and 2 led to vastly different state-dependent densities. Further, the ODBA values had many extreme values that needed to be accommodated, which further increased the difficulties of selecting appropriate state-dependent distributions. As ODBA is not a metric that can easily be divided into active/inactive behaviours in sharks, we estimated the state-dependent densities nonparametrically, in both states, in order to minimize the bias introduced by assigning inadequate parametric distributions \citep{lan15}. Figure \ref{fig:statedepdens} displays the fitted distributions.

To examine potential diel and tide effects on activity levels, we let the entries of the t.p.m.\ be functions of up to two covariates: time of day and tide level ({\it ebb, flood, low, and high}). Tide data was obtained from the NOAA tides and currents website for Palmyra Atoll and was processed by denoting high or low tide as $\pm$1 hour from reported high or low tide times. Time of day is represented by two trigonometric functions with period 24 hours, cos(2$\pi$t/86400) and sin(2$\pi$t/86400) (86,400 is the number of seconds in a day). We use three indicator variables, $x_{1t}, x_{2t}$ and $x_{3t}$, for tide levels \textit{high}, \textit{flood}, and \textit{ebb}, respectively, so that the entries of the t.p.m.\ have the following form 
\begin{equation*}
\text{logit}(\gamma_{ij}(t)) = \beta_{0i} + \beta_{1i}\text{sin}(2\pi t/86400) + \beta_{2i}\text{cos}(2\pi t/86400) + \beta_{3i}x_{1t} + \beta_{4i}x_{2t} + \beta_{5i}x_{3t}
\end{equation*}
for $i=1,2$, $j\neq i$, $t=1,\ldots, 86400$. The intercept term $\beta_{0,i}$ corresponds to \textit{low} tide. 

\begin{table}[!htb]
\centering
\begin{tabular}{l c c c c c}
\hline
Model & Log-likelihood & AIC & $\Delta$ AIC & BIC & $\Delta$ BIC \\
\hline \hline
No covariates & 639299.2& -1278370 & 779 & -1277178 & 692 \\
Time & 639558.1 & -1278872 & 277.2 & -1277645 & 225\\
Time, High & 639657.6 & -1279063 & 86.2 & -1277819 & 51\\
Time, High, Flood & 639695.2 & -1279130 & 19 & -1277869 & 1\\
\textbf{Time, High, Flood, Ebb} & 639708.7 & -1279149  & \textbf{0} &  -1277870 & \textbf{0}\\
\hline
\end{tabular}
\caption{Model fitting results for a blacktip shark. Based on the AIC and BIC, the model selected includes time of day and includes differences in activity levels based on all categories of tide levels.}
\label{tab:btresults}
\end{table}
 
\begin{figure}[!htb]
\centering
\includegraphics[scale=.85]{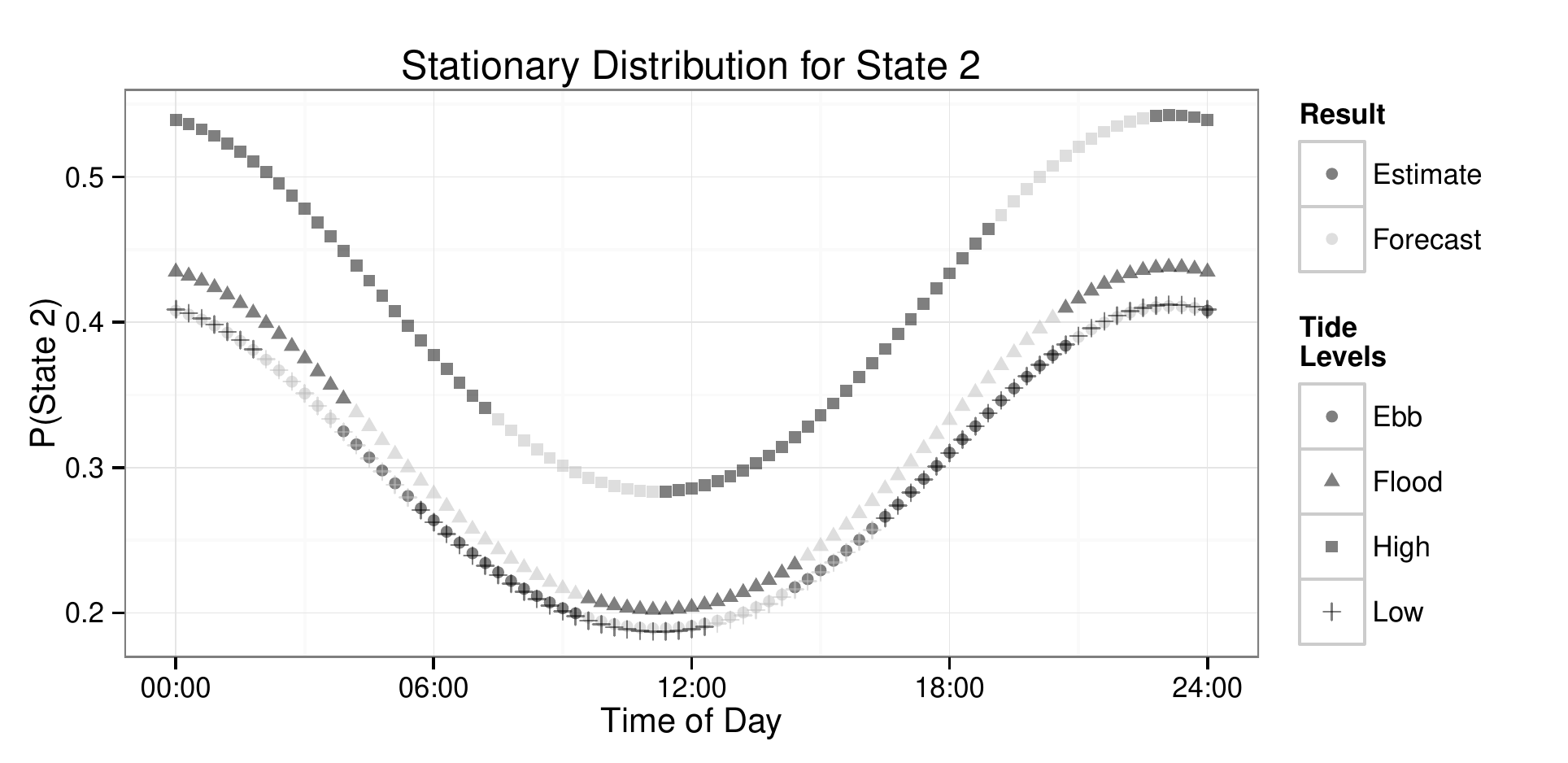}
\caption{Implied stationary distribution for state 2, the more active state, by time of day and tide level. For tide levels, we distinguish between model estimates, such that the corresponding tide level was observed at that time of day, and forecasts.}
\label{fig:btbstatdist}
\end{figure}

Based on the selected model (cf.\ Table \ref{tab:btresults}), the shark's activity levels were, on average, lowest from approximately 9:00 -- 13:00 and highest from 21:00 -- 1:00. In Figure \ref{fig:btbstatdist}, we see that the shark was more active during high tide in general when compared to flood, low or ebb tide. While the equilibrium (or stationary) distribution associated with low and ebb tide overlap, the state-dwell probabilities are higher during ebb tide than in low tide.

\begin{figure}[!htb]
\centering
\includegraphics[scale=.8]{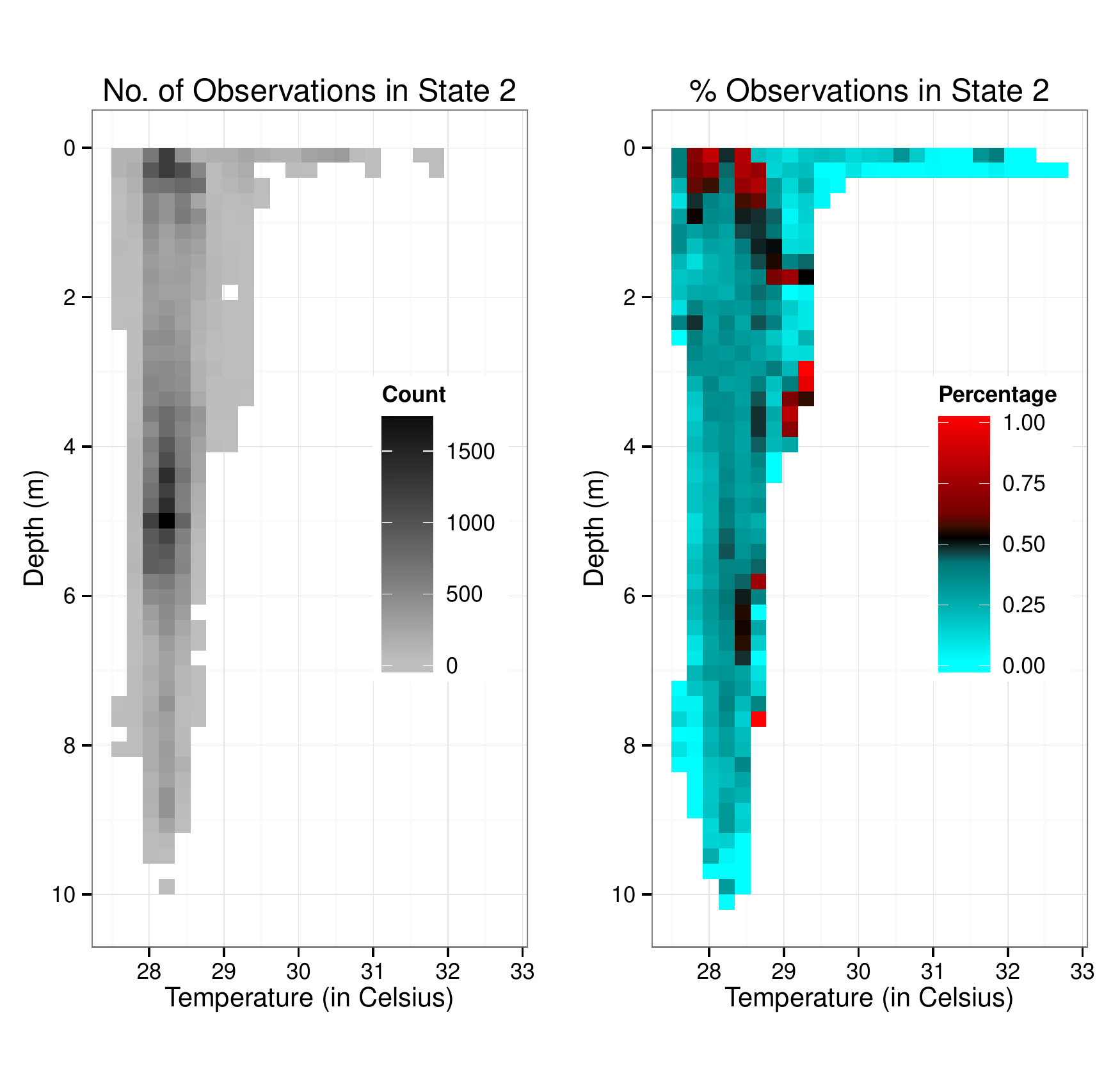}
\caption{The number of observations in each grid cell that correspond to state 2. Zero counts appear in white. (left) Percentage of observations in each cell that correspond to state 2. (right)}
\label{fig:depthtemp}
\end{figure}

Using the Viterbi algorithm, we decoded the optimal state sequence of the ODBA time series. To further understand the effect of vertical habitat on behaviour, we related the decoded state sequence results to a grid of depth and temperature values, shown in Figure \ref{fig:depthtemp}. The shark spent most of its time over the nearly five day period in depths of about 3-6 metres and between 28-29 degrees Celsius, with some higher counts also in shallower waters, which is reflected in the state 2 counts. However, the percentages of state 2 observations reveals that the shark was generally more active when near the surface in waters of 28-29 degrees C. There was generally less active behaviour exhibited when the individual was in very shallow warm water ($>$ 29 degrees C). 

\section{Discussion}
We considered two approaches for analysing animal accelerometer data with HMMs: a supervised learning approach, such that classification is of primary interest, and an unsupervised learning approach. The aim of a study and the type of data available will determine which of the two is to be preferred. When the objective is to do classification and there is a set of pre-defined behaviours of interest, then the model's ability to correctly predict and categorize behaviours is of main interest. In this instance, a supervised learning approach may be applied. One of the benefits of such an approach is that the behavioural classes are exactly defined, making interpretation relatively straightforward. Alternatively, if the objective is to infer (or, colloquially speaking, to `learn') new aspects of animal behaviour, then the unsupervised approach provides an excellent framework. The latter comes with the implicit caveat that the states will not necessarily map directly to specific animal behaviours. Any post-hoc behavioural interpretation of the estimated states is directly connected to the metric(s) used, and must draw from background biological knowledge of the species of interest. In many cases, behaviours such as foraging may not be exclusive to one state or another. Nonetheless, if the model is able to identify bouts of behaviour which consistently re-appear, then it is often likely that these signify something important in the animal's behavioural repertoire and are worthy of further investigation. 

Even when classification is the goal of an analysis, there are certainly practical scenarios which preclude the use of an HMM, e.g.\ if the training data do not reflect the transitions between behaviours or if there is insufficient data. Multiple studies have shown that other machine learning algorithms, e.g.\ support vector machines (SVM) or random forests, can work well for classification of animal acceleration data \citep{mar09, nat12, car14,gra15}. However, disregarding the serial dependence in the acceleration data usually is an unrealistic assumption, which often goes unmentioned or is treated as an afterthought. Adopting the assumption of independence is particularly risky if inferential statistics are applied to the output of say a machine learning algorithm. In these cases, secondarily applied statistical tests will implicitly assume that the machine learning categorizations contain more information content than is warranted, potentially leading to spurious results. This is not just a statistical nuance and can be a crucial point. Such tests are often applied as decision making tools to sort out ``what matters'' and setting the direction for much further research effort. Also, in assuming independence, one allows for classifications that may not be biologically realistic or must filter the classifications to properly identify a specific behaviour. For instance, \cite{car14} used a SVM where one of the primary interests was to identify prey handling/capture for penguins at sea. To confirm a prey capture event, they ruled that if the SVM classified three consecutive observations as prey-handling this counted as a true prey capture. In contrast, an HMM would have bypassed the need to filter through the classification results by accounting for the serial dependence in observations corresponding to prey handling. Further, the ability to identify general behaviours when the data are processed over short time periods, such as a few seconds or less, will complicate the ability of any machine learning algorithm that assumes independence to properly classify a sequence of observations into the same class, unless the boundaries between classes are well-defined which may not always be the case. Many behaviours persist over longer stretches of time than those at which the data is processed, also necessitating the use of a model that can account for the serial dependence. 

In the literature, inference on behavioural state-switching dynamics has sometimes been made using two-stage (or even three-stage) analyses, where HMMs (or other machine learning algorithms) are used to decode the behaviours underlying given observations, and subsequently a logistic regression is conducted for relating the decoded behaviours to covariates (see, e.g., \citealp{har10,bro14}). The appeal of such an approach lies in the ease of implementation: fairly basic HMMs, without covariates, are fitted to the accelerometer data and used to decode the states, and, subsequently, standard regression software packages can be used to conduct a regression of the behavioural states on covariates. However, it is our view that such a multi-stage analysis is less suited to relating accelerometer data to covariates than the {\em joint} modelling approach presented in Section \ref{ss:covinf}, for two reasons: (i) in the multi-stage analyses, the uncertainty in state estimates is usually not propagated through the different stages of analysis, and (ii) a regression analysis on decoded states needs to take into account the high serial correlation in those states. Rather than ignoring these issues or trying to address them within a multi-stage analysis (which will render such an approach technically challenging), a direct joint modelling approach, where neither of the problems arise, seems preferable.

Using a direct joint modelling approach in Section \ref{sec:dataexam} we were able to learn about the effects that atmospheric variables have on activity levels of a soaring raptor, while for the blacktip reef shark we examined temporal and tidal inputs effects on its activity levels. The HMM produced similar temporal patterns of activity to a previous analysis of the blacktip reef shark data set using GAMMs \citep{pap15}.  Both analytical methods revealed crepuscular and/or nocturnal increases in activity with a tidal component, with the shark most active at the high tide or as tide was about to ebb. By incorporating swimming depth and temperature, it was also revealed that highest activity was seen when the shark was at the surface in waters of 28-29 C. More importantly, the analysis showed that the shark was inactive when in very warm ($>$29 C) shallow water or deeper water. These results agree with a previous hypothesis that sharks are `hunting warm, and resting warmer' and use warmer water ($>$ 29 C) to increase the rate of some physiological function such as digestion, and not for foraging \citep[see][]{pap15}.  The HMM in this case allows us to explain the drivers of activity in the shark and move beyond just describing its movements, but rather explain `why' it may be moving or selecting certain habitats. The HMM also provided a measure of the change in probability of the individual being in active states. Although there was a clear temporal pattern of activity, the HMM identified the shark as 30 percent more likely to be in an active state during the late evening hours. A probabilistic approach also makes it possible to compare patterns of  behaviours between sympatric species and how they may be differently influenced by temporal or environmental characteristics. Quantifying changes in probability between sympatric species will be of great importance when looking at competitive interactions or resource partitioning.

We have covered the basic HMM framework here, but the popularity of the HMM framework is due in part to its many extensions. In particular, there are two HMM extensions that have been proven useful in classification of human activities: the hidden semi-Markov model (HSMM) \citep{lan12b} and the hierarchical hidden Markov model (HHMM) \citep{fin98}. The HSMM models the time spent within a state by some probability distribution with support on the positive real integers, thereby allowing for more complex state dwell time distributions than can be provided by an HMM (namely only geometric distributions). For instance, an HMM may not model the time spent in a resting behaviour adequately if the animal is known to rest for long periods of time. The HHMM provides the framework necessary to identify composite behaviours. For instance, lunge feeding in baleen whales is a composite behaviour made up of (1) initial increase in acceleration with (2) a positive pitch angle, as animals commonly approach prey schools from below, followed by (3) a rapid deceleration after the whale opens its mouth increasing its drag \citep{owe15}.
The HHMM models each composite behaviour as its own HMM, and models the transitions between composite behaviours, i.e.\ switches between HMMs.

\section{Acknowledgements}

TP was supported by a South African National Research Foundation Scarce Skills postdoctoral research fellowship, and TP and YPP received funding from the MASTS pooling initiative (The Marine Alliance for Science and Technology for Scotland) and their support is gratefully acknowledged. MASTS is funded by the Scottish Funding Council (grant reference HR09011) and contributing institutions. TP and MM gratefully acknowledge the hardware, software, support and expertise contributed by Prof Willem Bouten and his research group UvA-BiTS (University of Amsterdam Bird Tracking System).

\section{Data Accessibility}

We expect to archive data from this manuscript in a public archive, such as Dryad. 

\renewcommand\refname{References}
\makeatletter
\renewcommand\@biblabel[1]{}

\end{spacing}
 
\end{document}